# Plädoyer für den Äther[*]

**Norbert Feist** [*]
D 86368 Gersthofen, Leo - Fall - Str. 26

**Zusammenfassung**. In einer Neubewertung des Michelson - Morley - Experimentes wird dargelegt, daß die Phasengeschwindigkeit $c_{(\varphi)}$ in einem innerhalb eines Mediums mit der Geschwindigkeit v bewegten Inertialsystem zu

$$c_{(\varphi)} = \frac{c^2 - v^2}{c + v \cos \varphi}$$

gemessen wird mit c als konstanter Phasengeschwindigkeit im ruhenden Medium und $\varphi$ als Winkel zwischen der Bewegungsrichtung des Systems und der Ausbreitungsrichtung der Wellen. Bei Bewegung einer Quelle wird die „Kreiswelle" zur Ellipse mit der numerischen Exzentrizität $\varepsilon = \frac{v}{c}$ mit Quelle und Beobachter im Brennpunkt. Beim Licht wie beim Schall ist nicht $c_{(\varphi)}$ isotrop und damit konstant sondern das harmonische Mittel $\overline{c_h}$ aus Hin- und Rückgeschwindigkeit, das sich wie folgt errechnet:

$$\overline{c_h} = \frac{c^2 - v^2}{c}$$

Diese richtungsunabhängige Durchschnittsgeschwindigkeit erklärt zwanglos den Ausgang des Experimentes. Ein wohl erstmals tatsächlich durchgeführtes Michelson - analoges Experiment mit Schallwellen zeigt in Übereinstimmung mit obigen Gleichungen im Gegensatz zu bisherigen Annahmen das gleiche Ergebnis wie das Michelson – Morley – Experiment mit Licht. Da sich die Existenz der Luft als Trägermedium für den Schall nicht leugnen läßt, spricht der Umkehrschluß für das notwendige Vorhandensein eines Äthers als Trägermedium für elektromagnetische Wellen. Deshalb wird abschließend versucht, auch die der SRT zugeschriebene Geschwindigkeitsabhängigkeit von Masse und Zeit (=Frequenz) als mechanische Begleitumstände einer jeden Beschleunigung quantitativ darzustellen.

**Schlagwörter**. Lichtgeschwindigkeit, Michelson-Morley-Experiment, Spezielle Relativitätstheorie

## 1. Definitionen

Äther = Hypothetisches Ausbreitungsmedium für elektromagnetische Wellen
SRT = Spezielle Relativitätstheorie
$c$ = Licht- oder Schallgeschwindigkeit, im ruhenden Medium in jeder Richtung konstant, d.h. isotrop.
$v$ = Gleichförmige Geschwindigkeit eines allseits offenen Inertialsystems gegenüber dem ruhenden Medium (In der SRT : Relativgeschwindigkeit gegenüber einem anderen Inertialsystem).
$\varphi$ = Winkel zwischen der Bewegungsrichtung des Systems und der Ausbreitungsrichtung der Wellen aus Sicht des Systems.
$c_{(\varphi)}$ = Licht- oder Schallgeschwindigkeit gemessen im Inertialsystem. Gemäß Äthertheorie auch bei Licht abhängig von $\varphi$ und $v$, in der SRT immer gleich $c$.
$\overline{c_h}$ = Im Inertialsystem gemessener harmonischer Mittelwert aus Hin- und Rückgeschwindigkeit von Licht oder Schall in der Form $\frac{2}{\overline{c_h}} = \frac{1}{c_{(\varphi)}} + \frac{1}{c_{(\varphi + \pi)}}$, in der SRT gleich $c$.
Inertialsystem = Geradlinig und gleichförmig bewegtes System, in dem das Trägheitsgesetz gilt.

---

## 2. Die Lichtgeschwindigkeit in einem mit der Geschwindigkeit v bewegten Inertialsystem

### 2.1 Historisches

In Analogie zu Luft und Schall sah die Äthertheorie des 19. Jahrhunderts in einem im Raum ruhenden Äther das Ausbreitungsmedium für elektromagnetische Wellen. Infolge deren isotroper Phasengeschwindigkeit $c$ galt er gleichzeitig als das absolute Bezugssystem gegenüber anderen Geschwindigkeiten. Die einfache Vektoraddition von $c$ und $v$ ergibt aus Sicht eines innerhalb des Mediums bewegten Inertialsystems für jeden Winkel $\varphi$ eine andere Geschwindigkeit $c(\varphi)$. So schien es möglich, wie beim Schall durch Messung dieser Lichtgeschwindigkeit $c(\varphi)$ einmal in Bewegungsrichtung und einmal entgegengesetzt dazu die Geschwindigkeit $v$ des Inertialsystems Erde und die Geschwindigkeit $c$ des Lichtes selbst berechnen zu können. Nahm man für die Erde deren Bahngeschwindigkeit von 30 km/s an, so lag der Unterschied zwischen $c+v$ und $c-v$ jedoch innerhalb der damaligen Meßungenauigkeit und eine Absolutmessung schied allein deshalb schon aus. Michelson glaubte, mit seinem sehr genauen Interferometer Unterschiede zwischen der Hin- und Rückgeschwindigkeit des Lichtes in Bahnrichtung der Erde und senkrecht dazu erfassen und durch Drehen der Apparatur in Form von Interferenzstreifenverschiebungen sichtbar und damit meßbar machen zu können. Als harmonischen Mittelwert der Geschwindigkeiten in und entgegen der Fahrtrichtung nahm er richtigerweise an

$$\frac{2}{\overline{ch}(0/180)} = \frac{1}{c-v} + \frac{1}{c+v} \tag{1}$$

$$\overline{ch}(0/180) = \frac{c^2 - v^2}{c} = c\left(1 - \frac{v^2}{c^2}\right) \tag{2}$$

Bei seinem ersten Versuch 1881 glaubte er, daß sich das Licht - aus Sicht des Systems - senkrecht zur Fahrtrichtung und zurück völlig unbeeinflußt mit $c$ ausbreiten würde. Beim folgenden Versuch 1887 zusammen mit Morley wurde hierfür gemäß der Vektorrechnung der klassischen Kinematik mit

$$c(\varphi) = \sqrt{c^2 - v^2 \sin^2 \varphi} - v \cos \varphi \tag{3}$$

$$\overline{Ch} = \frac{c^2 - v^2}{\sqrt{c^2 - v^2 \sin^2 \varphi}} \tag{4}$$

oder in diesem Falle bei 90 Grad auch einfach nach Pythagoras angenommen

$$c(90) = \overline{Ch}(90/270) = \sqrt{c^2 - v^2} = c\sqrt{1 - \frac{v^2}{c^2}} \tag{5}$$

Aber, egal welchen Unterschied man zu Grunde legte und erwartete: Die Messung zeigte keine Streifenverschiebung. Es breitete sich Unbehagen aus, und es wurde z.B. überlegt, ob die Apparatur in Bewegungsrichtung durch den Ätherwind immer genau passend kontrahiert wird ( Fitzgerald, Lorentz ) oder ob sich die Geschwindigkeit der Quelle doch zur Lichtgeschwindigkeit hinzuaddiert (Ritz). Auf Grund der Einsteinschen Arbeit von 1905 [1] ging man schließlich davon aus, daß die Lichtgeschwindigkeit in jedem Inertialsystem unabhängig von dessen oder des Beobachters Geschwindigkeit immer und in allen Richtungen gleich $c$ sei und die elektromagnetischen Wellen im Unterschied zu allen anderen zur Ausbreitung kein Medium benötigen würden. Anders als beim Schall sollte es z.B. beim Licht ähnlich dem Relativitätsprinzip der klassischen Mechanik unmöglich sein, aus der Messung der Lichtgeschwindigkeit in einem System auf dessen Geschwindigkeit gegenüber dem Medium zu schließen. Bei einem Teil der Ätheranhänger wurde dadurch das Unbehagen jedoch nicht beseitigt sondern verstärkt. Bis in die heutige Zeit kritisieren Physiker wie Laien die SRT [z.B. 2,3]. Mit Sicherheit sind inzwischen viele Ätheranhänger gestorben mit dem unausgesprochenen Satz auf den Lippen „Und es gibt ihn doch".

## 2.2 Neubewertung des Michelson - Morley - Experimentes

Zweifelsohne haben neuere und genauere Wiederholungen des Experimentes das Ergebnis von Michelson bestätigt. Dieses lautete für ein bewegtes Inertialsystem jedoch nicht: „Die Lichtgeschwindigkeit ist in allen Richtungen gleich $c$ ". Das M-M - Experiment besagte lediglich: „Die harmonischen Mittelwerte $\overline{c_h}$ aus Hin- und Rückgeschwindigkeit der beiden betrachteten Lichtstrahlen sind gleich groß".

Als Vertreter der Äthertheorie hat man daraus folgenden Schluß zu ziehen: Die einfach zu überblickenden Verhältnisse in „Fahrtrichtung" mit $c_{(0°)} = c - v$ und $c_{(180°)} = c + v$ ergeben für den harmonischen Mittelwert entspr. Gl. (2) $\overline{c_{h(0/180)}} = \frac{c^2 - v^2}{c}$ . Das Ergebnis des Experimentes legt nahe, daß dieser Wert eine richtungsunabhängige, also isotrope Systemkonstante ist. Die Annahme einer transversalen mittleren Geschwindigkeit von $\overline{c_{h(90)}} = c$ (Jahr 1881) oder von $\overline{c_{h(90/270)}} = c\sqrt{1 - \frac{v^2}{c^2}}$ (Jahr 1887) entspr. Gl. (5) war dann falsch. Die falsche Annahme läßt vermuten, daß ohne Kenntnis der Detailvorgänge die üblicherweise für Massenpunkte in der Kinematik angewendete Vektoraddition versagt. Eine darauf ausgerichtete zeichnerische Analyse der zeitlichen Abläufe sollte zunächst wie für 0 Grad auch für 90 Grad und schließlich für jeden Winkel Wertepaare für $c_{(\varphi)}$ und $c_{(\varphi + \pi)}$ ergeben, deren harmonischer Mittelwert der Bedingung $\overline{c_h} = \frac{c^2 - v^2}{c}$ genügt.

Das Ergebnis dieser unter 2.4 dargestellten Analyse sei hier vorweggenommen: Aus Sicht von Quelle und mitbewegtem Beobachter beträgt in einem gegenüber dem ruhenden dispersionsfreien Medium mit der konstanten Geschwindigkeit $v$ bewegten offenen Inertialsystem gemäß Gl. (37) die winkelabhängige Größe der Lichtgeschwindigkeit:

$$c_{(\varphi)} = \frac{c^2 - v^2}{c + v \cos \varphi}$$

Das harmonische Mittel aus Hin- und Rückgeschwindigkeit ist gemäß Gl. (50) in allen Richtungen gleich groß und damit für jedes System eine richtungsunabhängige d.h. isotrope systemspezifische Konstante:

$$\overline{c_h} = \frac{c^2 - v^2}{c}$$

Neben der zeichnerischen Analyse der zeitlichen Abläufe wird dieses Ergebnis nicht nur gestützt durch das Michelson - Morley - Experiment und seine zahlreichen Wiederholungen:
Die Gleichungen (37) und (50) beschränken sich nämlich nicht nur auf das Licht. Sie sollten z.B. genauso für den Luftschall gelten. Seit fast 100 Jahren gibt es jedoch zwei Welten: Die Schallausbreitung sei als Gegenstand der Kontinuumsmechanik auf die Punktmechanik zurückführbar [4]. Im Gegensatz zu dem der SRT gehorchenden mediumlosen Licht könne man daher mit Michelson - analogen Schallexperimenten die Systemgeschwindigkeit gegenüber dem Medium Luft feststellen [z.B. 4,5,6]. Ein eigenes - nicht nur in Gedanken durchgeführtes - unter 3. beschriebenes Michelson - analoges Experiment mit Schallwellen ergab im Gegensatz zu dieser Lehrmeinung jedoch wie beim Michelson - Morley - Experiment keinen Unterschied zwischen den beiden Richtungen bezüglich der harmonischen Mittelwerte aus Hin- und Rückgeschwindigkeit und bestätigt die Gültigkeit der Gl. (37) und (50) auch für die Schallausbreitung.
Die neue zwanglose und quantitative Erklärung des Michelson - Morley - Experimentes und die jetzt im Ergebnis gleichartige Wellenausbreitung von Licht und Schall sprechen somit für die Existenz eines Licht - Äthers.

## 2.3 Vektor - Kurz - Darstellung dreier Theorien zu $c_{(\varphi)}$ und $\overline{c_h}$

Die Lichtgeschwindigkeit $c$ sei 2,5 cm/s und die Geschwindigkeit $v$ des Inertialsystems sei 1,5 cm/s. Als sich Quelle und Beobachter vor einer Sekunde in M befanden, wurden nach allen Richtungen hin ebene Wellen emittiert. B bezeichnet die derzeitige Position von Quelle und Beobachter. Länge und Richtung der Pfeile zeigen die Größe der Lichtgeschwindigkeit $c_{(\varphi)}$ an, die sich aus Sicht des mitbewegten Beobachters in der jeweiligen Richtung ergibt.

A   Äthertheorie bis zum Jahr 1905

Sie galt bis 1887 für Licht und analog für den Schall.
- Trägermedien für Licht und Schall erforderlich (Äther bzw. Luft z.B.)
- $c_{(\varphi)}$ = anisotrop entspr. Gl. (3)
- $\overline{c_h}$ = anisotrop entspr. Gl. (4)
- Raum und Zeit sind absolut

Sie konnte durch das Michelson - Morley - Experiment für Licht nicht bestätigt werden. Die Gültigkeit der Gleichungen für den Schall wurde aber bis heute weiterhin vorausgesetzt.

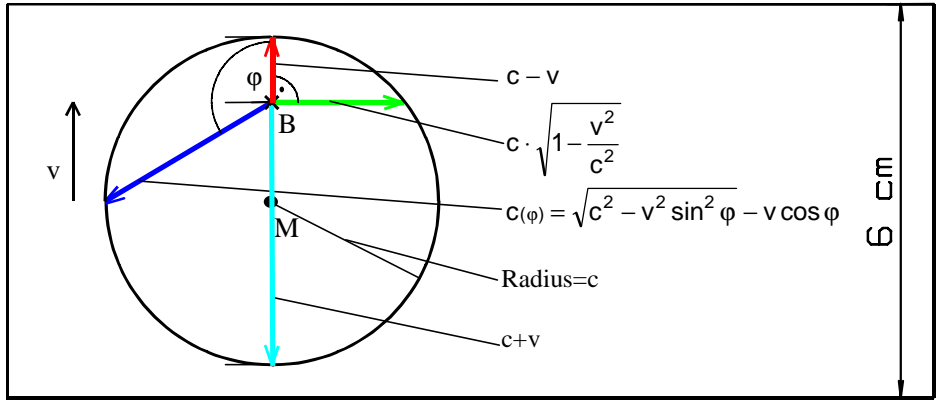

Abb. 1   Äthertheorie bis 1905

B   Spezielle Relativitätstheorie ab 1905

Sie gilt nur für die Ausbreitung elektromagnetischer Wellen und erklärt das Ergebnis des Michelson - Morley - Experimentes auf spezielle Weise. Sie ist auf die Ausbreitung von Schall- und anderen Wellen nicht anwendbar.
- Kein Trägermedium erforderlich
- $c_{(\varphi)} = \overline{c_h} = c$ = isotrop
- Weder $c_{(\varphi)}$ noch $\overline{c_h}$ sondern Weg- und Zeitmessung hängen von $v$ ab. Dadurch soll es weder einen absoluten Raum noch eine absolute Zeit geben.

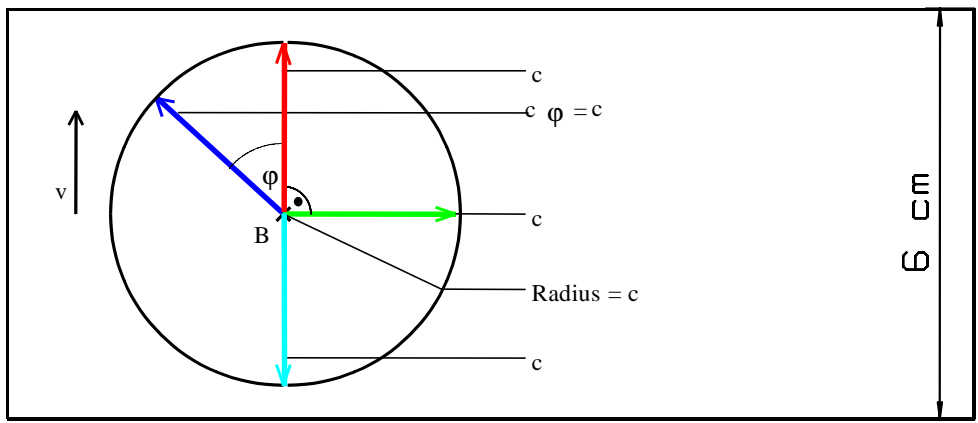

Abb.2  Spezielle Relativitätstheorie

## C  Äthertheorie gemäß dieser Arbeit

Die Gleichungen (37) und (50) gelten für Licht und analog unter anderem auch für Luftschall. Die Bestätigung erfolgte bisher durch das Michelson - Morley - Experiment und ein Michelson - analoges eigenes Experiment mit Schallwellen.

- Trägermedium Äther bzw. Luft erforderlich
- $c_{(\varphi)}$ = anisotrop entspr. Gl. (37). Beobachter und Quelle befinden sich bezüglich der Geschwindigkeitsvektoren in einem Brennpunkt einer Ellipse der numerischen Exzentrizität $\varepsilon = \dfrac{v}{c}$ mit c als großer Halbachse und v als linearer Exzentrizität
- $\overline{c_h} = \dfrac{c^2 - v^2}{c}$ =isotrop entspr. Gl.(50)
- Raum und Zeit sind (wieder) absolut.

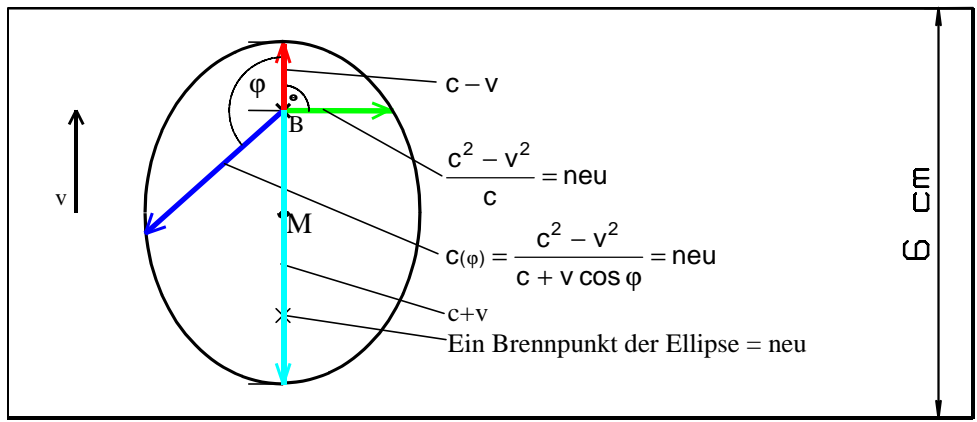

Abb. 3   Äthertheorie gemäß dieser Arbeit

## 2.4  Herleitung der Gleichungen für $c_{(\varphi)}$ und $\overline{c_h}$

Die Vorstellung von Kreiswellen ist nützlich bei der integralen Betrachtungsweise einer allseits emittierenden Punktquelle, z.B. zur Veranschaulichung der isotropen Ausbreitung oder zur Ermittlung der Flächenenergie beim quadratischen Abstandsgesetz.

Die differenzierte Betrachtung der energetischen Verhältnisse zeigte jedoch bereits beim Photoeffekt, daß die Kreiswelle als Individuum an sich nicht existiert. Einsteins „Nadelstrahlung" kam der Realität deutlich näher.

Eine differenzierte Betrachtung in der geometrischen oder Strahlenoptik erfordert bei einer punktförmigen ruhenden Emissionsquelle für jede dieser „Nadeln" die Annahme einer Tangente am Kreis als ebene Welle. Dieses Bild erlaubt die Anwendung des erfolgreichen Konzepts der „ebenen Welle" für jede Einzelnadel schon bei geringstem Abstand von der Quelle.

Wie im Folgenden zunächst am bewegten Spiegel gezeigt, ist die Neigung dieser ebenen Wellen gegenüber der Emissionsrichtung wie die Flugrichtung eines emittierten Massenpunktes im absoluten Raum dem Trägheitsgesetz folgend von der Geschwindigkeit der Quelle abhängig. Diese Neigung der Wellenebene führt im bewegten System zu einer gegenüber der klassischen Vorstellung abweichenden emissionswinkelabhängigen Geschwindigkeit $c_\varphi$ und vor allem zu einem vom Winkel unabhängigen Mittel aus Hin- und Rückgeschwindigkeit. Die Neigung der Wellennormale ist das Geheimnis des Michelson - Morley - Experimentes.

### 2.4.1  Transversale Reflexion am 45 Grad - Spiegel des Interferometers

Für die zeichnerische Analyse der zeitlichen Abläufe werden gewählt $c = 3\,cm/s$ und $v = 1\,cm/s$. Die Frequenz sei $\nu = 1/s$ und die Wellenlänge $\lambda = 3\,cm$.

Abbildung 4 zeigt die Reflexion einer ebenen Welle am ruhenden 45 Grad - Spiegel.

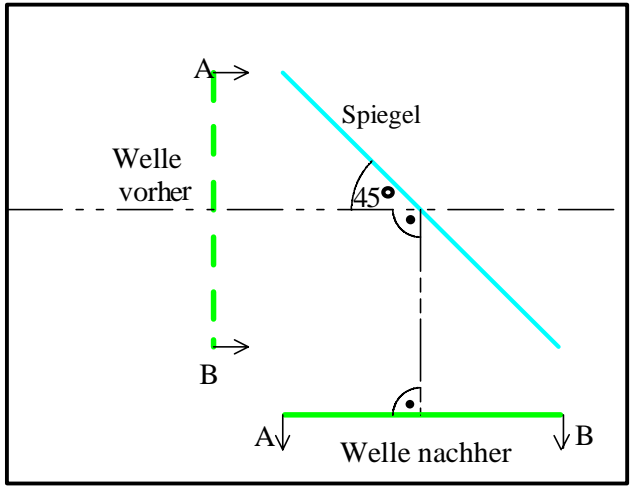

Abb. 4   Reflexion einer ebenen Welle am ruhenden 45 Grad - Spiegel

Die Abbildung 5 gibt die Reflexion einer ebenen Welle am mit $v = 1\,\text{cm/s}$ bewegten 45 Grad- Spiegel wieder. $t$ sec nach Auftreffen des Punktes A der Welle trifft auch Punkt B den Spiegel. $t$ berechnet sich für ein $l$ von 4 cm gemäß

$$c \cdot t = l + vt \qquad (6)$$

zu $t = \dfrac{l}{c - v} = 2\,\text{sec}$. Aus $c \cdot t = 6\,\text{cm}$ und $v \cdot t = 2\,\text{cm}$ ergeben sich die Orte A und B der reflektierten Welle. Die Wellennormale - hier definiert als Senkrechte auf der ebenen Welle - neigt sich nach Änderung der Emissionsrichtung um 90 Grad um den Winkel $\beta$ in Fahrtrichtung mit $\tan\beta = \dfrac{v}{c}$. Die Geschwindigkeitskomponente in transversaler Richtung ist weiterhin $c$.

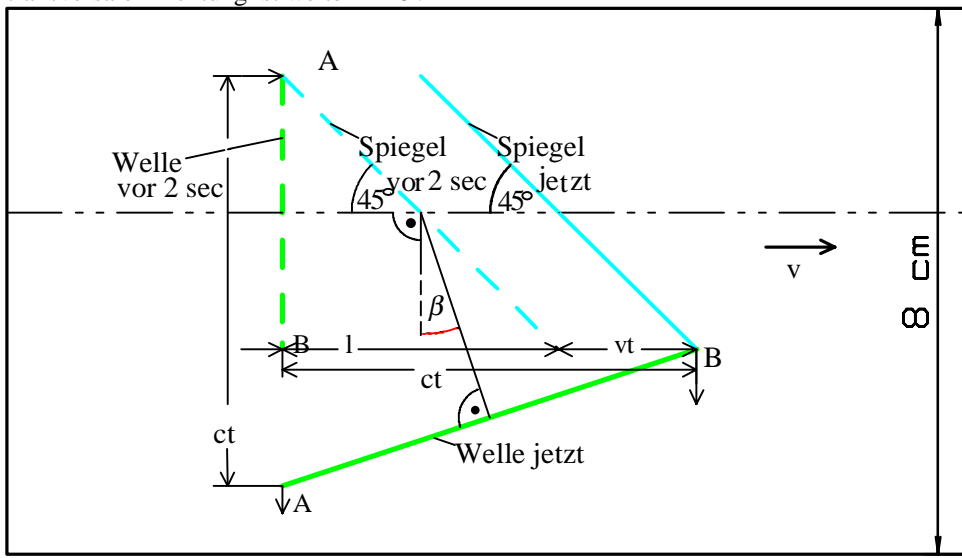

Abb. 5   Reflexion einer ebenen Welle am mit $v$ bewegten Spiegel.

Die Abbildung 6 gibt die Situation von Abb. 5 wieder, ergänzt um die 1 Sekunde zuvor reflektierte Welle, deren Punkt B in der gestrichelt gezeichneten Situation den Spiegel 1 cm vorher traf und die inzwischen 3 cm nach „unten" gewandert ist:

Abb. 6   Situation wie in Abb. 5, ergänzt um die eine Sekunde vorher reflektierte 1. Welle.

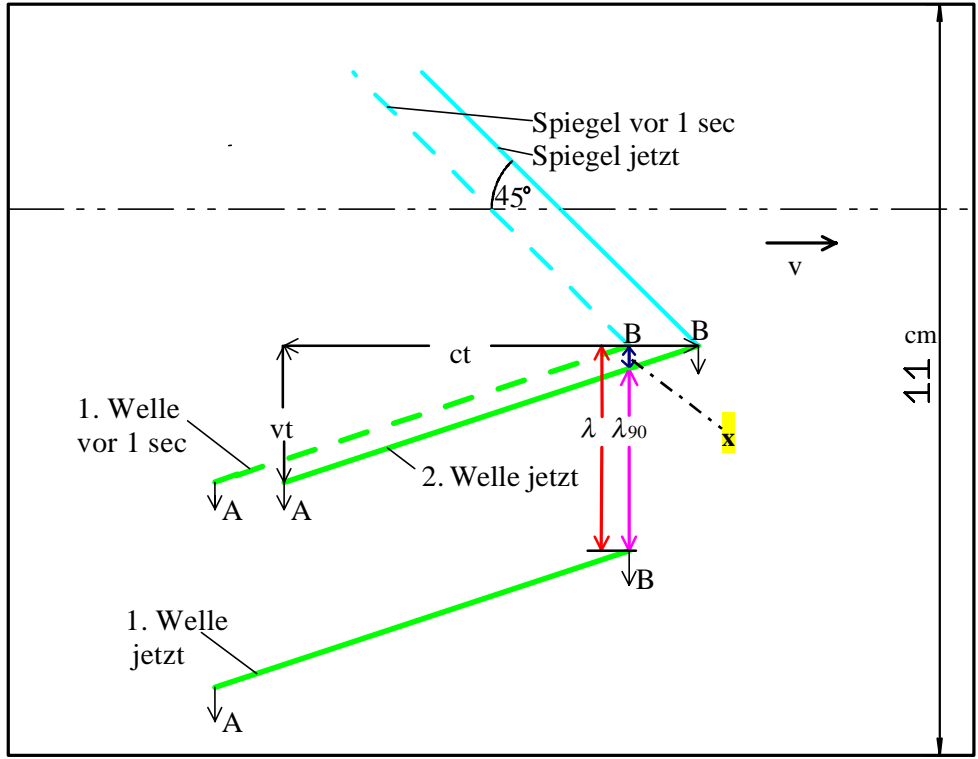

Die waagerechte Strecke $\overline{BB}$, um die der Spiegel vorrückte, ist gleich $\frac{v}{\nu}$. Nach dem Strahlensatz ist

$$\frac{x}{v \cdot t} = \frac{\frac{v}{\nu}}{c \cdot t} \tag{7}$$

$$x = \frac{v^2}{c \cdot \nu} \tag{8}$$

Mit $\nu = \frac{c}{\lambda}$ wird daraus

$$x = \lambda \cdot \frac{v^2}{c^2} \tag{9}$$

Die transversale Wellenlänge $\lambda_{90}$ beträgt dann

$$\lambda_{90} = \lambda - x \tag{10}$$

$$\lambda_{90} = \lambda - \lambda \cdot \frac{v^2}{c^2} = \lambda\left(1 - \frac{v^2}{c^2}\right) \tag{11}$$

In diesem Beispiel sind das

$$\lambda_{90} = 3\,\text{cm}\left(1 - \frac{1^2}{3^2}\right) = 2{,}67\,\text{cm}$$

Hier soll jedoch die transversale Lichtgeschwindigkeit $c_{90}$ ermittelt werden. Aus der für ein im Äther ruhenden System gültigen Gleichung

$$c = \nu \cdot \lambda \tag{12}$$

wird im mit v bewegten System

$$c_{(\varphi)} = \nu \cdot \lambda_\varphi \tag{13}$$

bzw. im transversalen Fall

$$c_{(90)} = \nu \cdot \lambda_{90} \tag{14}$$

$$c_{(90)} = \nu \cdot \lambda \left(1 - \frac{v^2}{c^2}\right) = c\left(1 - \frac{v^2}{c^2}\right) = \frac{c^2 - v^2}{c} \tag{15}$$

Dieses Ergebnis ist auch zeichnerisch nachvollziehbar an Hand der Abbildung 7.

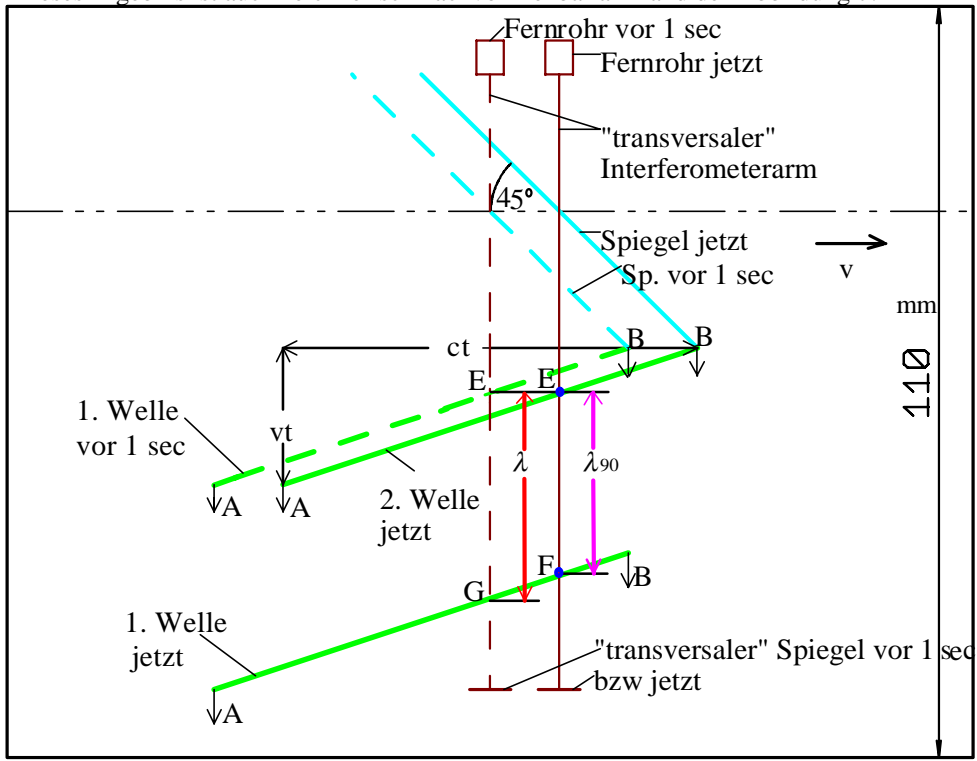

Abb. 7  Die Situation von Abb. 6 ist um den transversalen Interferometerarm mit Fernrohr und Spiegel ergänzt worden.

Die gestrichelt gezeichnete Welle, die vor einer Sekunde den Interferometerarm bei E geschnitten hat, hat sich in dieser einen Sekunde im Raum um λ von E nach G bewegt. Auf dem sich gleichzeitig nach rechts bewegenden Arm ist sie in der einen Sekunde für den mitbewegten Beobachter meßbar jedoch nur um λ₉₀ von E nach F gewandert. Zu diesem Zeitpunkt schneidet bei E bereits die nächste Welle den Arm. Auf dem zur Messung benutzten Interferometerarm beträgt daher für den mitbewegten Beobachter die transversale Lichtgeschwindigkeit entsprechend Gl. (15)

$$c_{(90)} = \nu \cdot \lambda_{90} = c\left(1 - \frac{v^2}{c^2}\right) = \frac{c^2 - v^2}{c}$$

Nach Reflexion am transversalen Spiegel ergeben sich spiegelbildlich die gleichen Verhältnisse für den Rückweg. Das harmonische Mittel aus beiden Geschwindigkeiten ist ebenso groß und damit gleich dem des longitudinalen Interferometerarmes, was den Ausgang des Michelson-Morley-Experimentes erklärt.

**2.4.2 Transversale Emission**

Für die zeichnerische Analyse der zeitlichen Abläufe werden gewählt $c = 12\,\mathrm{cm/s}$ und $v = 4\,\mathrm{cm/s}$. Die Frequenz sei wieder $\nu = 1/\mathrm{s}$.

Die Verhältnisse bei transversaler Emission sind zwar analog zur Reflexion, jedoch ergibt sich der Winkel β zwischen der oben definierten Wellennormalen und der Emissionsrichtung nicht automatisch durch die Zeichnung. Man muß auf die Erfahrung gestützt verschiedene Annahmen machen. Den Gesetzen der Trägheit gehorchend nimmt ein in einem mit v bewegten Inertialsystem transversal mit w geworfener Ball für einen nicht mitbewegten Beobachter beispielsweise die in Abb. 8 dargestellte Richtung β und Geschwindigkeit u an.

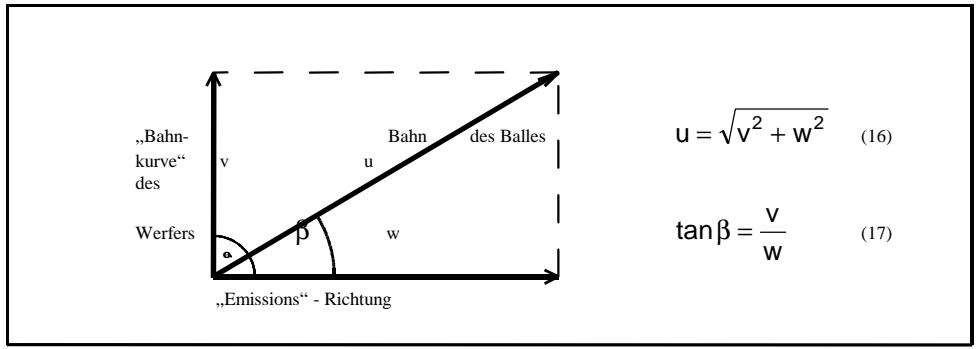

Abb. 8  Bahnkurve eines Massenpunktes (Ball) aus Sicht eines nicht mitbewegten Betrachters. Der transversal „emittierte" Ball bleibt immer auf Höhe des selbst mit v bewegten Werfers.

Licht besitzt erwiesenermaßen ebenfalls eine Masse. Außerdem zeigt es bei Versuchen auf der Erde keine Aberrationserscheinungen. Kann man es als Außenstehender, nicht mit der Quelle mitbewegter Experimentator in Form der transversal emittierten Bremsstrahlung, der Synchrotonstrahlung oder bei der Zerstrahlung beobachten, so sieht man die Vorwärtsrichtung im Raum, qualitativ der Abb. 8 entsprechend. Quantitativ sieht es geringfügig anders aus: Das Licht gehorcht zwar den Trägheitsgesetzen, da aber $w$ in diesem Fall gleich der Grenzgeschwindigkeit $c$ ist, kann $u$ nicht größer als $c$ sein. Wellen können nicht schneller sein als die sie tragenden oder umgebenden Teilchen, eine Nachricht kann nicht schneller sein als der Bote. Beim Schall ist die Geschwindigkeit der Wellen von der Lufttemperatur und damit von der Geschwindigkeit der Gasmoleküle abhängig, und das wird beim Äther nicht anders sein, siehe Punkt 4. Es ist anzunehmen, daß die Wellennormale wie beim Ball um den Winkel $\beta$ mit $\tan\beta = \dfrac{v}{c}$ nach vorn „klappt" mit der Maßgabe, daß die senkrechte Geschwindigkeitskomponente gleich $c$ bleibt, siehe Abb. 9.

Abb. 9  Transversale Emission aus Sicht eines nicht mitbewegten Betrachters

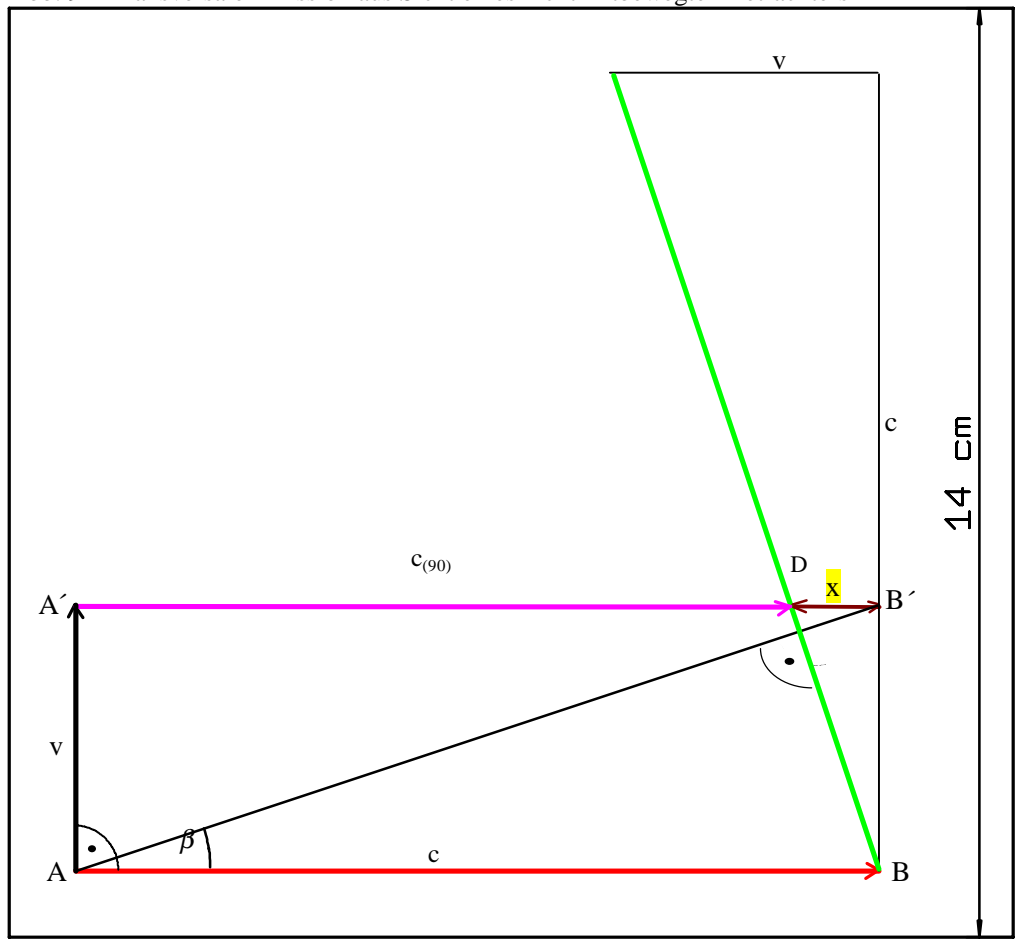

Im absoluten Raum hat die ebene Welle auf der Strecke AB nach einer Sekunde den Punkt B erreicht. Für den mitbewegten Beobachter schneidet die ebene Welle die Strecke A´B´ zu diesem Zeitpunkt im Punkt D, so daß er die Geschwindigkeit

$$c_{(90)} = c - x \tag{18}$$

mißt. Nach dem Strahlensatz gilt

$$\frac{x}{v} = \frac{v}{c} \tag{19}$$

$$x = \frac{v^2}{c} \tag{20}$$

Mithin hat die transversale Lichtgeschwindigkeit im mit v bewegten System auch im primären Emissionsfall wie bei der zuvor behandelten Reflexion entsprechend Gl. (15) die Größe

$$c_{(90)} = c - \frac{v^2}{c} = \frac{c^2 - v^2}{c} = c\left(1 - \frac{v^2}{c^2}\right)$$

### 2.4.3 Emission unter beliebigem Winkel φ gegenüber v

Abb. 10   Emission unter beliebigem Winkel φ

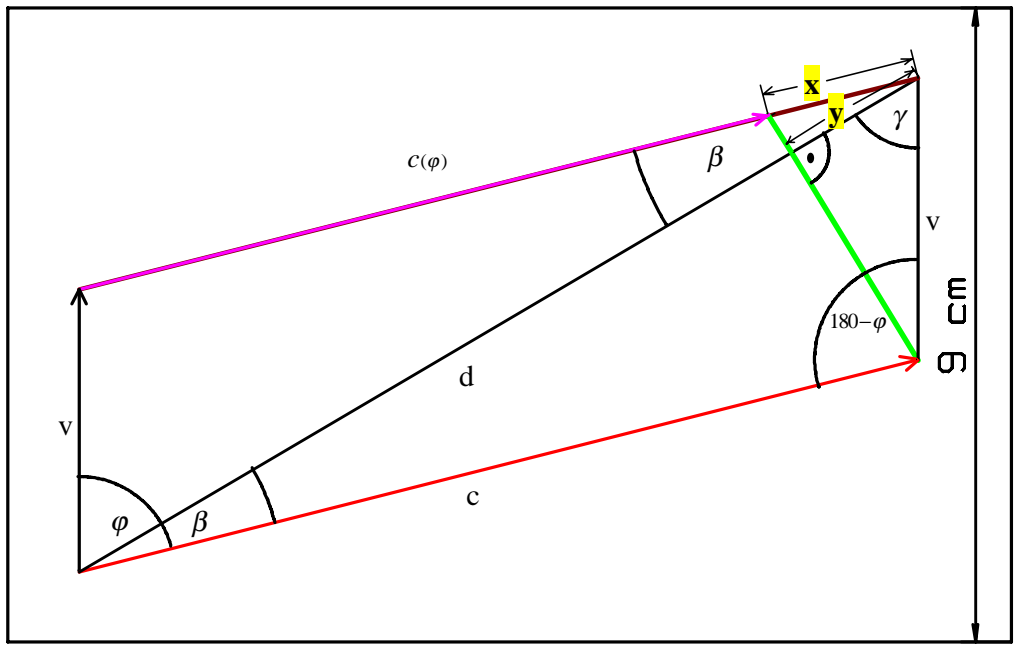

In der Abb. 10 berechnet sich die Diagonale d des Parallelogramms nach dem Kosinussatz

$$d^2 = v^2 + c^2 - 2vc\cos(180 - \varphi) \tag{21}$$

bzw. wegen

$$\cos(180 - \varphi) = -\cos\varphi \tag{22}$$

$$d^2 = v^2 + c^2 + 2vc\cos\varphi \tag{23}$$

$$d = \sqrt{v^2 + c^2 + 2vc\cos\varphi} \tag{24}$$

Des weiteren besagt der Kosinussatz

$$c^2 = v^2 + d^2 - 2vd\cos\gamma \tag{25}$$

Daraus folgt mit

$$\cos\gamma = \frac{y}{v} \tag{26}$$

$$\cos \gamma = \frac{v^2 + d^2 - c^2}{2vd} = \frac{y}{v} \tag{27}$$

$$y = \frac{v^2 + d^2 - c^2}{2d} \tag{28}$$

Gemäß Kosinussatz ist ferner

$$v^2 = c^2 + d^2 - 2cd \cos \beta \tag{29}$$

Dies ergibt zusammen mit

$$\cos \beta = \frac{y}{x} \tag{30}$$

$$\frac{c^2 + d^2 - v^2}{2cd} = \frac{y}{x} \tag{31}$$

Einsetzen der Werte für $y$ und $d$ ergibt

$$x = \frac{(v^2 + d^2 - c^2)c}{c^2 + d^2 - v^2} = \frac{(v^2 + v^2 + c^2 + 2vc \cos \varphi - c^2)c}{c^2 - v^2 + v^2 + c^2 + 2vc \cos \varphi} = \frac{(2v^2 + 2vc \cos \varphi)c}{2c^2 + 2vc \cos \varphi} \tag{32}$$

$$x = c \frac{v^2 + vc \cos \varphi}{c^2 + vc \cos \varphi} \tag{33}$$

Es folgt mit

$$c_{(\varphi)} = c - x \tag{34}$$

$$c_{(\varphi)} = c - c \frac{v^2 + vc \cos \varphi}{c^2 + vc \cos \varphi} = c - \frac{v(v + c \cos \varphi)}{c + v \cos \varphi} \tag{35}$$

$$c_{(\varphi)} = \frac{c(c + v \cos \varphi) - v(v + c \cos \varphi)}{c + v \cos \varphi} = \frac{c^2 + vc \cos \varphi - v^2 - vc \cos \varphi}{c + v \cos \varphi} \tag{36}$$

Hieraus wird die allgemeine Gleichung (37) für die Lichtgeschwindigkeit in einem mit $v$ bewegten System mit $\varphi$ als Winkel zwischen Bewegungs- und Emissionsrichtung:

$$c_{(\varphi)} = \frac{c^2 - v^2}{c + v \cos \varphi} \tag{37}$$

Sie ist vom Typus her eine Kegelschnittgleichung in Parameterdarstellung.

Die Prüfung für die beim Michelson - Experiment aus Sicht des bewegten Systems relevante transversale und longitudinale Richtung ergibt nach Einsetzen der Kosinuswerte folgendes:

- Für den transversalen Arm des Michelson - Interferometers resultiert mit jeweils $\cos \varphi = 0$ für $\varphi = 90$ und für $\varphi = 270$ Grad für den Hinweg wie für den Rückweg

$$c_{(90/270)} = \frac{c^2 - v^2}{c} \tag{38}$$

- Das harmonische Mittel der auf Hin- und Rückweg gleich großen Geschwindigkeiten beträgt ebenfalls

$$\overline{c_{h(90/270)}} = \frac{c^2 - v^2}{c} . \tag{39}$$

- Auf dem „longitudinalen" Arm gilt für den Hinweg mit $c^2 - v^2 = (c+v)(c-v)$ und $\cos 0° = 1$ der seit langem bekannte Wert

$$c_{(0)} = \frac{(c+v)(c-v)}{c+v} = c - v \tag{40}$$

und für den Rückweg mit $\cos 180° = -1$ ebenso bekanntermaßen

$$c_{(180)} = \frac{(c+v)(c-v)}{c-v} = c + v \tag{41}$$

Das harmonische Mittel beider Geschwindigkeiten beträgt entspr. Gl. (2) wie oben

$$c_{h(0/180)} = \frac{c^2 - v^2}{c} \tag{42}$$

Diese Gleichheit beider mittlerer Geschwindigkeiten würde zunächst nur das Michelsonergebnis für den üblichen Fall der rechtwinklig angeordneten Interferometerarme erklären.

Es soll nun ferner überprüft werden, ob dieses harmonische Mittel aus Hin- und Rückgeschwindigkeit, wie zu Beginn des Punktes 2.2 aus dem Ergebnis des Michelson - Versuchs gefolgert, bei konstanter Geschwindigkeit v eine isotrope Systemkonstante ist und für jedes beliebige Winkelpaar aus $\varphi$ für den Hinweg und $(\varphi + \pi)$ für den Rückweg gilt.

Das harmonische Mittel $\overline{c_h}$ aus den jeweils entgegengesetzt gerichteten Lichtgeschwindigkeiten $c_\varphi$ und $c_{(\varphi + \pi)}$ lautet in allgemeiner Form:

$$\frac{2}{\overline{c_h}} = \frac{1}{c_\varphi} + \frac{1}{c_{(\varphi + \pi)}} \tag{43}$$

$$\overline{c_h} = \frac{2 c_{(\varphi)} \cdot c_{(\varphi + \pi)}}{c_{(\varphi)} + c_{(\varphi + \pi)}} \tag{44}$$

Wegen

$$\cos(\varphi + \pi) = -\cos\varphi \tag{45}$$

erhält die nach Gl. (37) einzusetzende Geschwindigkeit für den Rückweg unter Beibehaltung des Winkelausdruckes $\varphi$ lediglich ein Minuszeichen im Nenner:

$$c_{(\varphi + \pi)} = \frac{c^2 - v^2}{c - v \cos\varphi} \tag{46}$$

Durch Einsetzen beider Geschwindigkeiten wird dann

$$\overline{c_h} = \frac{\dfrac{2(c^2 - v^2)(c^2 - v^2)}{(c + v \cos\varphi)(c - v \cos\varphi)}}{\dfrac{c^2 - v^2}{c + v \cos\varphi} + \dfrac{c^2 - v^2}{c - v \cos\varphi}} \tag{47}$$

$$\overline{c_h} = \frac{\dfrac{2(c^2 - v^2)(c^2 - v^2)}{(c + v \cos\varphi)(c - v \cos\varphi)}}{\dfrac{(c^2 - v^2)(c - v \cos\varphi) + (c^2 - v^2)(c + v \cos\varphi)}{(c + v \cos\varphi)(c - v \cos\varphi)}} \tag{48}$$

$$\overline{c_h} = \frac{2(c^2 - v^2)(c^2 - v^2)}{(c^2 - v^2)(c - v \cos\varphi + c + v \cos\varphi)} = \frac{2(c^2 - v^2)}{2c} \tag{49}$$

Das ergibt $$\overline{c_h} = \frac{c^2 - v^2}{c} \tag{50}$$

was zu beweisen war. Der harmonische Mittelwert aus Hin- und Rückgeschwindigkeit in einem mit v bewegten Inertialsystem ist isotrop, das heißt in jeder Richtung gleich groß. Für das Michelson - Interferometer bedeutet dies, daß - auch für den Fall ungleicher Armlängen oder anderer Armwinkel - die Anzahl der Wellenlängen pro Arm bei Drehung der Apparatur richtungsunabhängig konstant bleibt. Daher kann es zu keiner „Streifenverschiebung" kommen.

Als mathematischer Nebeneffekt ergibt sich übrigens durch die Analogie von Geschwindigkeitsvektoren und Brennstrahlen eine **allgemeinere Definition** für den Parameter einer Ellipse. Dieser ist nicht nur gleich der halben zur Hauptachse senkrechten Sehne durch den Brennpunkt sondern gleich dem harmonischen Mittel aus den beiden Brennstrahlen jeder beliebigen Sehne durch den Brennpunkt.

## 3. Ergebnisse von Schallmessungen

Nachdem wie dargelegt das Michelson - Ergebnis nicht durch eine Voigt - Lorentz - Transformation erklärt werden muß, war zu erwarten, daß die Gleichungen (37) und (50) auch auf die Ausbreitung von Schallwellen im Medium Luft anzuwenden sind. Insbesondere sollte dann - anders als bisher allgemein angenommen - bei einem Michelson - analogen Versuch mit Schallsignalen bei mit konstanter Geschwindigkeit bewegter Sende- und Empfangsvorrichtung kein richtungsabhängiger Laufzeitunterschied der Echos feststellbar sein. Anders ausgedrückt: Der harmonische Mittelwert aus Hin- und Rückgeschwindigkeit der Schallwellen sollte in Fahrtrichtung und senkrecht dazu und auch in jeder anderen Richtung gleich sein und der Gleichung (50) genügen:

$$\overline{c_h} = \frac{c^2 - v^2}{c}$$

Zur Bestätigung wurde der Wert von $\overline{c_h}$ experimentell bei verschiedenen Winkeln und Geschwindigkeiten ermittelt.

### 3.1 Meßprinzip und verwendete Geräte

Zur Messung dienten ein Ultraschall - Abstandsmeßgerät und eine Meßstrecke bekannter Länge auf einem PKW. Die Ultraschallabstandsmessung beruht auf dem Puls/Echo - Verfahren, bei dem aus der gemessenen Laufzeit eines Schallimpulses zum reflektierenden Objekt und zurück bei vorgegebener mittlerer Schallgeschwindigkeit die Entfernung zum Objekt berechnet wird. Ist wie in diesem Fall die Entfernung bekannt, kann umgekehrt aus dem sich bei Bewegung des Systems ändernden Entfernungsmeßwert der jeweils zugehörige harmonische Mittelwert aus Hin- und Rückgeschwindigkeit des Schalls errechnet werden.

Eine aus Ultraschallwandler und Reflektor bestehende horizontal drehbare Meßstrecke war auf dem Dachgepäckträger einer Limousine montiert. Im Winkelbereich 0 bis 68 Grad zur Fahrtrichtung kam ein Abstandsmeßgerät LR3S Typ 262 der Fa. Format Messtechnik mit 50 kHz - Folienwandler zum Einsatz.. Bei 90 Grad wurde das Gerät P42-M3A-2D-1G1-2205 der Fa. PIL Ultraschallsensorik mit keramischem 220 kHz - Wandler und doppeltem Meßbereich verwendet. In der Meßstrecke befand sich der Fühler eines NiCr-NiAl - Thermoelementes mit einer Zeitkonstante von 25 ms. Bei Windstille und möglichst gleichmäßiger Temperatur wurden bei verschiedenen Winkeleinstellungen zur Fahrtrichtung jede Sekunde mittels Datenrecorder im zeitlichen Abstand von 125 ms die Werte von Länge, PKW - Geschwindigkeit und Temperatur abgefragt und gespeichert.

### 3.2 Auswertung

Es seien die Schallgeschwindigkeit in Luft bei $20°C$ $c = 343{,}37 m/s$ und die Geschwindigkeit des PKW $v = 100 km/h$ oder $v = 27{,}78 m/s$. Die Länge der Meßstrecke bei $20°C$ und $v = 0 km/h$ sei - auch mit dem Lineal gemessen - $l_0 = 1350 mm$.

Nach Gl. (50) $\overline{c_h} = c - \frac{v^2}{c}$ ist bei 100 km/h und $20°C$ - egal in welchem Winkel zur Fahrtrichtung - mit folgender mittleren Schallgeschwindigkeit $\overline{c_{h(100)}}$ für Hin- und Rückweg zu rechnen:

$$\overline{c_{h(100)}} = 343{,}37 - \frac{27{,}78^2}{343{,}37} = 341{,}12 m/s \qquad (51)$$

Die mittlere Schallgeschwindigkeit wird also kleiner und damit die Signallaufzeit größer. Der Prozessor des Abstandsmeßgerätes rechnet aber weiterhin mit den einprogrammierten $c = 343{,}37 m/s$ und interpretiert die längere Laufzeit bei der Geschwindigkeit v als größeren Abstand $l_{(v)}$ gegenüber der Länge $l_0$ bei Stillstand. Es gilt die Relation

$$l_{(v)} = \frac{l_0 \cdot c}{\overline{c_h}} \qquad (52)$$

$$l_{(100)} = \frac{1350 \cdot 343{,}37}{341{,}12} = 1358{,}7 \qquad (53)$$

Bei $v = 100 km/h$ sollte das Abstandsmeßgerät also bei einer 1350 mm - Meßstrecke 8,7 mm mehr anzeigen als im Stillstand. Die Standardabweichung der Messung von $l_0$ im Stillstand wurde zu 0,6 mm bestimmt.

Stellt man Gl. (52) um,

$$\overline{c_h} = c \frac{l_0}{l_{(v)}} \tag{54}$$

und setzt $c = 100\%$, so lautet die zur Ergebnisdarstellung benutzte Gleichung

$$\overline{c_h} = 100 \frac{l_0}{l_{(v)}} \; / \; \% \tag{55}$$

Die Schallgeschwindigkeit in Luft erhöht sich - wie mit der Meßstrecke im Stillstand gesondert überprüft - im relevanten Meßbereich mit der Temperatur um 0,18 %/K. An Stelle des üblicherweise in genaueren Entfernungsmeßgeräten eingebauten Temperaturfühlers zur automatischen elektronischen Anpassung des Schallgeschwindigkeitswertes an die Meßtemperatur wurde ein notwendigerweise schneller ansprechendes externes Temperaturmeßsystem verwendet. Die registrierten Ultraschallentfernungsmeßwerte $l_0$ im Stillstand und $l_v$ bei der Geschwindigkeit v wurden daher vor Anwendung der Gl. (55) entsprechend dem jeweils zugehörigen ebenfalls registrierten Temperaturmeßwert $\vartheta$ mit dem Faktor

$$k = \left[1 - (20 - \vartheta) \cdot 0{,}0018\right] \tag{56}$$

multipliziert und so auf eine vergleichbare Temperatur von $20\,°\mathrm{C}$ normiert.

**5.2 Meßergebnisse**

Die aus den Meßdaten gewonnenen harmonischen Mittel $\overline{c_h}$ aus Hin- und Rückgeschwindigkeit bezogen auf die Ruhe - Schallgeschwindigkeit $c$ werden in den Abbildungen 11 bis 15 in Relation zur Geschwindigkeit v der Schallquelle für verschiedene Emissionswinkel $\varphi$ graphisch dargestellt.

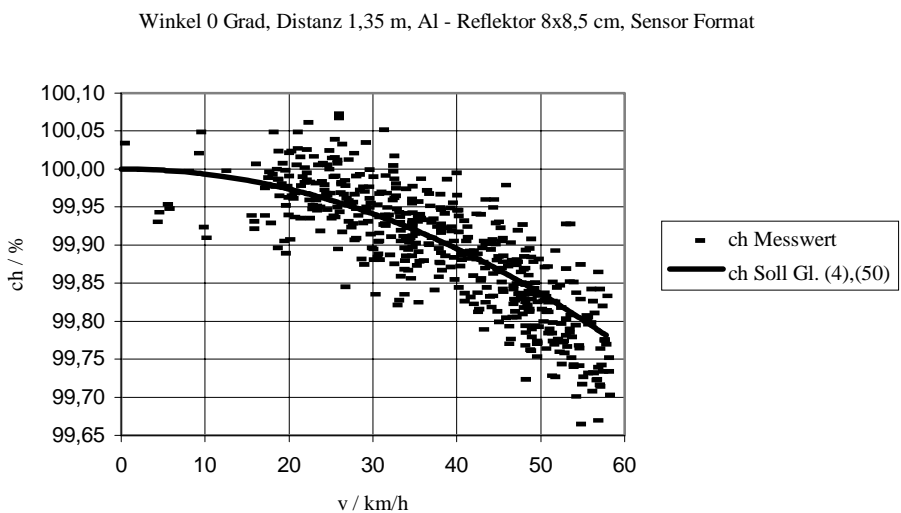

Abb. 11 Harmonisches Mittel $\overline{c_h}$ der Schallgeschwindigkeit für den Winkel $\varphi = 0$ Grad. Die durchgezogene Sollkurve entspricht der winkelunabhängigen Gl. (50) dieser Arbeit und hier im Falle von 0 Grad auch der Gl. (4) gemäß Lehrmeinung.

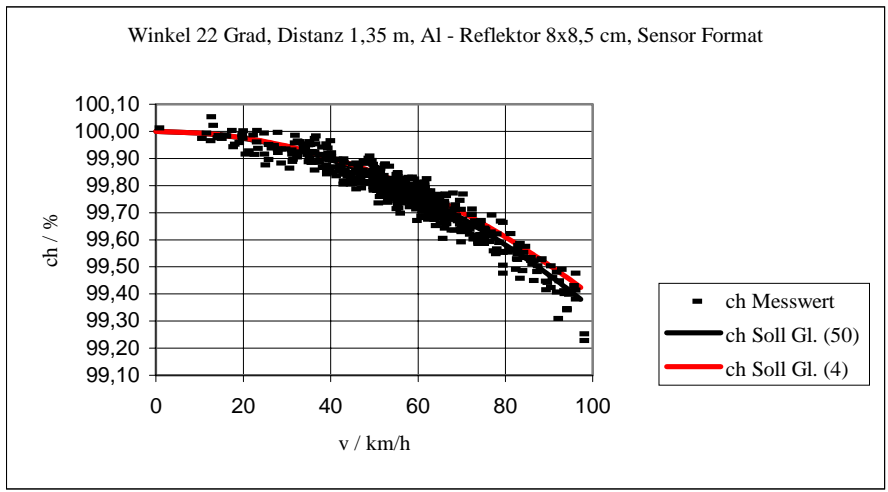

Abb. 12 Harmonisches Mittel $\overline{c}_h$ der Schallgeschwindigkeit für den Winkel $\varphi = 22$ Grad. Die schwarze Kurve entspricht der winkelunabhängigen Gl. (50) dieser Arbeit. Die rote Kurve wäre die Sollkurve gemäß Lehrmeinung entsprechend Gl. (4).

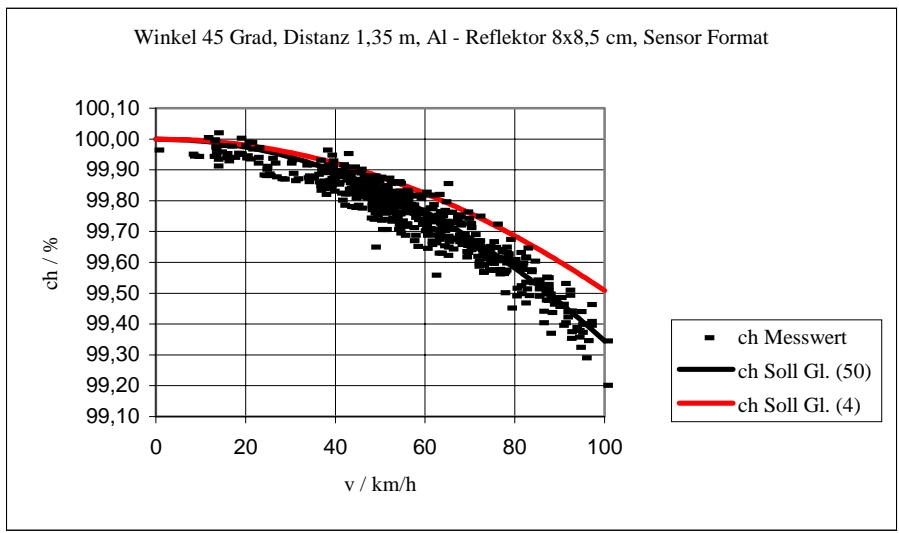

Abb. 13 Harmonisches Mittel $\overline{c}_h$ der Schallgeschwindigkeit für den Winkel $\varphi = 45$ Grad. Die schwarze Kurve entspricht der Gl.(50) dieser Arbeit. Die rote Kurve wäre die Sollkurve gemäß Lehrmeinung entsprechend Gl.(4).

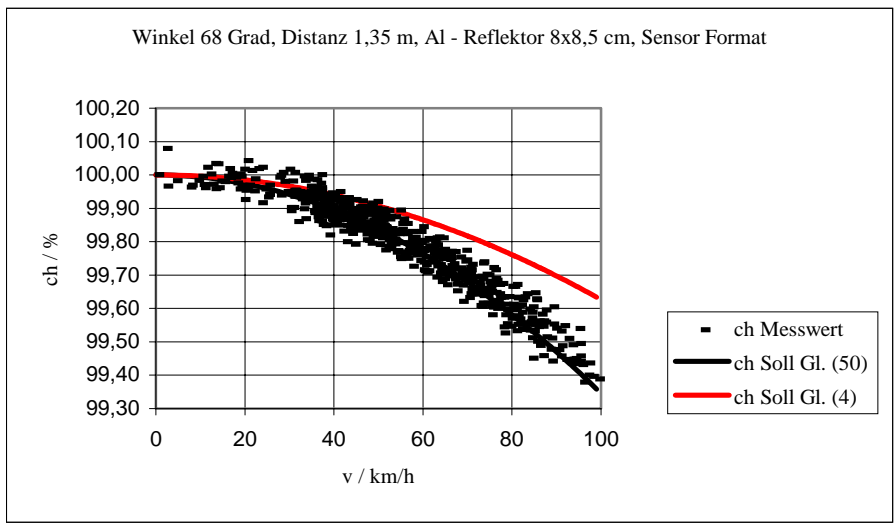

Abb. 14 Harmonisches Mittel $\overline{c}_h$ der Schallgeschwindigkeit für den Winkel $\varphi = 68$ Grad. Die schwarze Kurve entspricht der winkelunabhängigen Gl. (50) dieser Arbeit. Die rote Kurve wäre die Sollkurve gemäß Lehrmeinung entsprechend Gl.(4).

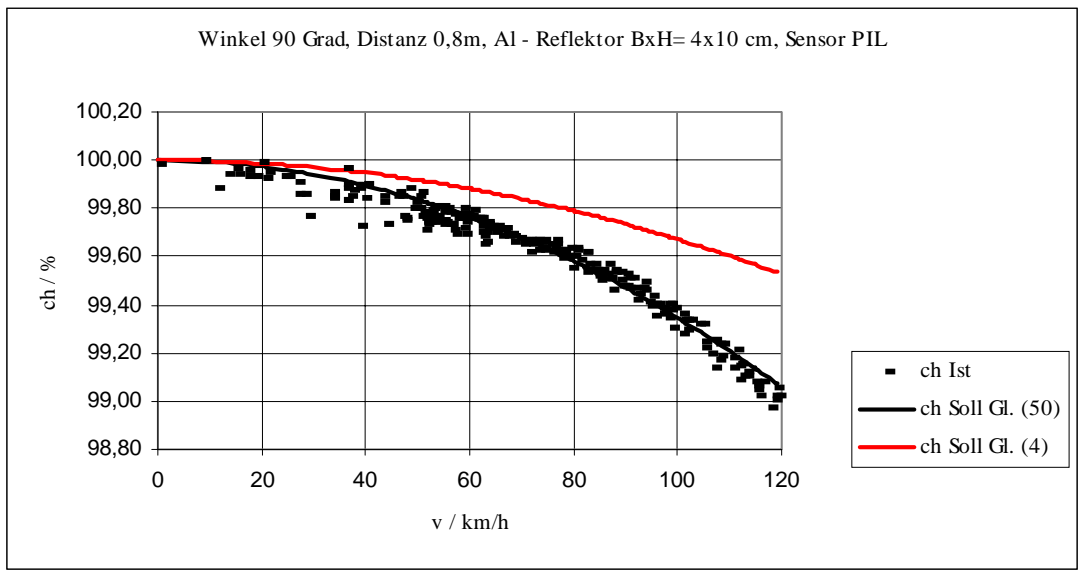

Abb. 15  Harmonisches Mittel $\overline{c_h}$  der Schallgeschwindigkeit für den Winkel  $\varphi = 90$ Grad. Die schwarze Kurve entspricht der winkelunabhängigen Gl. (50) dieser Arbeit. Die rote Kurve wäre die Sollkurve gemäß Lehrmeinung entsprechend Gl. (4).

### 3.4  Diskussion der Meßergebnisse

Die Meßwerte streuen jeweils um die schwarze „Soll" - Kurve entsprechend der in dieser Arbeit vorgestellten richtungsunabhängigen Gl. (50):  $\overline{c_h} = \dfrac{c^2 - v^2}{c} = c\left(1 - \dfrac{v^2}{c^2}\right)$, hier ebenfalls bezogen auf c = 100%.

Die Abbildungen 12 bis 15 enthalten außerdem zum Vergleich eine rote Kurve, basierend auf der offensichtlich falschen herkömmlichen Gleichung (4), ebenfalls in der hier üblichen Darstellung in Relation zu c =100%. Ausgehend vom Winkel 0 Grad, wo die Gl. (4) und (5) identische Werte ergeben, weichen die roten Sollkurven der winkelabhängigen Gl. (4) mit zunehmendem Winkel zunehmend von der Realität ab.

Im Rahmen der Meßgenauigkeit wird deutlich, daß in Übereinstimmung mit den entwickelten Gleichungen (37) und (50) der harmonische Mittelwert aus Hin- und Rückgeschwindigkeit bei konstanter Systemgeschwindigkeit v in allen Richtungen gleich groß ist. Dies und der Abschnitt 2.4 legen nahe, daß auch in der Optik die bisherige Vorstellung von dessen Winkelabhängigkeit als Basis des optischen Michelson - Morley - Experimentes zu korrigieren ist.

### 4.  Folgerungen

- Ein Plädoyer für den Äther schließt natürlich den Appell ein, die Lichtgeschwindigkeit aus dem Prokrustesbett der SRT zu befreien. Aus Gl. (37) und (50) folgt unter anderem, daß z.B. - vorbehaltlich einer Bestätigung durch die Zukunft - die bisher durch Laufzeitmessungen für Hin- und Rückweg ermittelten Werte für die Vakuumlichtgeschwindigkeit nicht $c$ sondern $\overline{c_h} = \dfrac{c^2 - v^2}{c}$ mit v als Geschwindigkeit der Erde im Äther zum Ergebnis hatten und daß sicher in einigen bewährten Formeln $c$ durch die Ausdrücke $c_{(\varphi)}$ bzw. $\overline{c_h}$ zu ersetzen ist.

Begleitend zu erneuten Untersuchungen zur Physik oder Thermodynamik des Äthers würden die in den letzten 100 Jahren in Mikro- und Makrokosmos gewonnenen und verworfenen Erkenntnisse neu zu bewerten sein.

- Bezüglich des Mikrokosmos ergeben sich Ansatzpunkte, wenn man die Lichtgeschwindigkeit als ganz normale Schallgeschwindigkeit des Äthers ansieht. Die Temperatur T des Äthers sollte derjenigen der Hintergrundstrahlung von 2,735 K entsprechen. Setzt man diese Temperatur unter der Annahme eines isothermen Vorgangs mit k als Boltzmannkonstante in die Gleichung für die Schallgeschwindigkeit = Lichtgeschwindigkeit

ein $$c = \sqrt{\dfrac{kT}{m}} \qquad (57)$$

so ergibt sich die Masse m der Ätherteilchen zu 4,2•10⁻⁴⁰kg. Das wäre der zweimilliardste Teil der Masse eines Elektrons.

- Bezüglich des Makrokosmos ergeben sich - abgesehen von der dunklen Äthermaterie - vor allem durch die Betrachtung der Lichtgeschwindigkeit als temperaturabhängige Schallgeschwindigkeit des Äthers Fragen zur zeitlichen und räumlichen Konstanz (nicht nur) der Lichtgeschwindigkeit. Postuliert man eine Abkühlung des Universums beispielsweise durch Verdampfen und nicht durch Expansion (möglich ist natürlich beides), so besagt z. B. die Rotverschiebung einer entfernten Galaxie von z = 4, daß die Lichtgeschwindigkeit damals dort fünf mal höher war als hier und heute. Bei gleicher Emissionsfrequenz wie heute wäre dann dort die Wellenlänge zwangsläufig bereits bei der Emission 5 mal größer als heute gewesen. Wegen der $\sqrt{T}$ - Abhängigkeit der Schall- oder Lichtgeschwindigkeit wäre dann als Ursache für die gemessene Rotverschiebung an Stelle einer Ausdehnung für den damaligen Emissionszeitpunkt eine 25 mal höhere Temperatur gegenüber heute anzunehmen. Das wären dann dort 68 K gewesen.

- Insbesondere schließen natürlich Spezielle Relativitätstheorie und absolute Äthertheorie einander aus. Nach dem hier vorgestellten Modell der anisotropen Lichtausbreitung in einem bewegten System ist die Annahme einer von Relativgeschwindigkeiten abhängigen Längenkontraktion nicht erforderlich. Sie ist im Gegensatz zu den anderen beiden in der SRT behandelten realen Effekten der Geschwindigkeitsabhängigkeit von Masse und „Zeit" auch bisher nicht experimentell nachgewiesen worden. In einer absoluten Äthertheorie bedarf die Geschwindigkeitsabhängigkeit von Masse und „Zeit" natürlich einer Begründung auf Basis von klassischer und Quantenmechanik. Im folgenden Abschnitt wird daher eine Prinzip - Skizze für die physikalischen Vorgänge bei Beschleunigungen entworfen unter der Annahme, daß Photonen - was immer man im einzelnen heute noch darunter verstehen mag - die Basis jeglicher Materie bzw. Masse oder „Ruhmasse" bilden, daß sie fast beliebig teilbar und in ihrer Gesamtheit unvergänglich sind und daß sie im Translationsfall als elektromagnetische Strahlung trotz der Möglichkeit zu zirkularer Polarisation vermutlich keinen Eigendrehimpuls oder Spin [7] haben.

## 5. Die Geschwindigkeitsabhängigkeit von Masse und Frequenz ( Prinzipskizze )

### 5.1 Definitionen:
$c$ = Vakuumlichtgeschwindigkeit
$m_{Ph}$ = Masse Photon
$m_E$ = Masse Elektron in Ruhe
$r$ = Photonenringradius des Elektrons
$\omega$ = Winkelgeschwindigkeit $2\pi s^{-1}$
$\nu$ = Kreisfrequenz = $\dfrac{\omega}{2\pi}$
$v_{rot}$ = Umfangsgeschwindigkeit = $r \cdot \omega$
$v_{trans}$ = Translationsgeschwindigkeit
$J$ = Trägheitsmoment = $m \cdot r^2$
$L$ = Drehimpuls = $J \cdot \omega$
$E_{rot}$ = Rotationsenergie = $\dfrac{1}{2} J \omega^2 = \dfrac{1}{2} m r^2 \cdot \dfrac{v_{rot}^2}{r^2} = \dfrac{1}{2} m v_{rot}^2$
$E_{trans}$ = Translationsenergie = $\dfrac{1}{2} m \cdot v_{trans}^2$

### 5.2 Prinzip:

Ein Elektron der Masse $m_E$ ruhe im Äther und bestehe aus Photonen der Summe 0,511 MeV, die mit der Umfangsgeschwindigkeit $v_{rot} = c$ um ein gemeinsames ( leeres ) Zentrum kreisen. Mit Elektronenmasse $m_E$ und Planckscher Wirkungskonstante h wird das Elektron so zu einem gedachten Punkt, der von einer Photonenhülle mit dem Radius und der Frequenz der bekannten Comptonwellenlänge des Elektrons umkreist wird.

Unter den Stabilitätskriterien, daß bei einem primären Materieteilchen Drehimpuls L und(!) Radius r konstant bleiben, kann das Elektron z.B. durch Stoß oder elektrisches Feld unter Einhaltung der Erhaltungssätze für Masse, Energie und Drehimpuls weitere Photonen aufnehmen. Da die Masse der aufgenommenen Photonen nicht(!) verschwindet, nimmt die Masse des Elektrons entsprechend zu:

$$m_E + m_{Ph} = m_{(E + Ph)} \qquad (58)$$

woraufhin sich diese Kombination unter Abnahme der Kreisfrequenz ν und der Umfangsgeschwindigkeit v$_{rot}$ mit der Translationsgeschwindigkeit v$_{trans}$ im Äther bewegen muß. Bei Energieabgabe läuft der umgekehrte Prozeß ab.

Im Folgenden wird quantitativ gezeigt, warum die Massenänderung eine Geschwindigkeits- und Frequenzänderung bewirkt. Bei der Aufnahme von Photonen durch ein Elektron lautet der Drehimpulserhaltungssatz

$$L_E + L_{Ph} = L_{(E + Ph)} \tag{59}$$

Wenn aber, wie angenommen, Photonen im Translationszustand keinen Drehimpuls mitführen, wohl aber nach der Absorption ihren Anteil am Bahndrehimpuls aller Photonen bzw. am Spin des Elektrons haben, wird daraus

$$L_E = L_{(E + Ph)} \tag{60}$$

$$J_E \cdot \omega_E = J_{(E + Ph)} \cdot \omega_{(E + Ph)} \tag{61}$$

$$\frac{1}{2} m_E \cdot r^2 \cdot \omega_E = \frac{1}{2} m_{(E + Ph)} \cdot r^2 \cdot \omega_{(E + Ph)} \tag{62}$$

Mit $\omega = 2\pi \cdot \nu$ resultiert daraus

$$m_E \cdot \nu_E = m_{(E + Ph)} \cdot \nu_{(E + Ph)} \tag{63}$$

$$\nu_{(E + Ph)} = \nu_E \frac{m_E}{m_{(E + Ph)}} \tag{64}$$

Gl. (64) drückt bereits die experimentell bekannte Tatsache aus, daß sich die Frequenzänderung umgekehrt proportional zur Massenänderung verhält.

Der Energieerhaltungssatz legt nahe, daß in einem abgeschlossenen System die Summe aus translatorischer und Rotationsenergie der Reaktionspartner Elektron und Photon konstant ist. Die Summe aus Rotationsenergie des ruhenden Elektrons und der Translationsenergie des Photons vor der Vereinigung ist gleich der Summe aus Rotations- und Translationsenergie der jetzt mit v$_{trans}$ bewegten Kombination aus beiden:

$$E_{rotE} + E_{transPh} = E_{rot(E+Ph)} + E_{trans(E+Ph)} \tag{65}$$

$$\frac{1}{2} m_E \cdot c^2 + \frac{1}{2} m_{Ph} \cdot c^2 = \frac{1}{2} m_{(E + Ph)} \cdot v_{rot}^2 + \frac{1}{2} m_{(E + Ph)} \cdot v_{trans} \tag{66}$$

$$m_{(E + Ph)} \cdot c^2 = m_{(E + Ph)} \cdot v_{rot}^2 + m_{(E + Ph)} \cdot v_{trans}^2 \tag{67}$$

$$c^2 = v_{rot}^2 + v_{trans}^2 \tag{68}$$

$$v_{rot} = \sqrt{c^2 - v_{trans}^2} \tag{69}$$

Durch die Aufnahme von Photonen = Masse verringert sich nicht nur die Kreisfrequenz sondern auch die Umfangs- oder Rotationsgeschwindigkeit der Photonen, die sich jetzt nicht mehr mit c sondern gleichzeitig mit v$_{rot}$ und v$_{trans}$ im absoluten Raum bewegen, wobei der Vektor der Translationsgeschwindigkeit senkrecht auf dem der Rotationsgeschwindigkeit steht, erkenntlich unter anderem an der Richtung der Dipolstrahlung.

Es ist
$$L = J \cdot \omega = m \cdot r^2 \cdot \frac{v_{rot}}{r} = m \cdot r \cdot v_{rot} \tag{70}$$

Es war ferner $L_E = L_{(E + Ph)}$

Daraus wird

$$m_E \cdot r \cdot c = m_{(E + Ph)} \cdot r \cdot v_{rot} \tag{71}$$

Und mit Gl.(69) ergibt sich

$$m_E \cdot c = m_{(E + Ph)} \cdot \sqrt{c^2 - v_{trans}^2} \tag{72}$$

Daraus folgt die bekannte Relation zwischen Masse und Translationsgeschwindigkeit

$$m_{(E + Ph)} = m_E \frac{c}{\sqrt{c^2 - v_{trans}^2}} = m_E \frac{1}{\frac{\sqrt{c^2 - v_{trans}^2}}{\sqrt{c^2}}} \tag{73}$$

$$m_{(E + Ph)} = \frac{m_E}{\sqrt{1 - \frac{v_{trans}^2}{c^2}}} \tag{74}$$

Dieser erstmals 1901 von Kaufmann bei der Beschleunigung von Elektronen beobachtete und von Einstein später mathematisch als Formel korrekt angegebene Zusammenhang zwischen Masse und Geschwindigkeit findet mit der Unvergänglichkeit der Photonen ( Erhaltung der Photonenmasse ) eine physikalisch einleuchtende Erklärung.

Die Relation zwischen Frequenz und Translationsgeschwindigkeit zeigt sich, wenn der Ausdruck für die Gesamtmasse $m_{(E + Ph)}$ aus Gl. (74) in die Gl. (64) eingesetzt wird

$$\nu_{(E + Ph)} = \nu_E \frac{m_E}{\frac{m_E}{\sqrt{1 - \frac{v_{trans}^2}{c^2}}}} \tag{75}$$

$$\nu_{(E + Ph)} = \nu_E \sqrt{1 - \frac{v_{trans}^2}{c^2}} \tag{76}$$

Diese mit jeder Beschleunigung verbundene ebenfalls mechanisch begründete Frequenzänderung - 1938 erstmals von Ives und Stilwell [8] als solche gemessen und gedeutet - wird in der SRT als sogen. Zeitdilatation interpretiert. Die Abhängigkeit der Taktdauer einer Atomuhr von deren Geschwindigkeit ist jedoch ähnlich banal wie die Abhängigkeit der Taktdauer einer Pendeluhr von deren Pendellänge und hat mit dem Fluß der Zeit an sich wenig zu tun.

**Literatur**

*) Anmerkung zum Photonenspin
Ebenso wie die SRT ist u. a. auch die Realität einer Quantisierung des elektromagnetischen Feldes anzuzweifeln. E. Hecht schreibt in seinem Lehrbuch „Optik" unter 13.3 „Der photoelektrische Effekt - Einsteins Photonenkonzept" in einer Fußnote sehr treffend: Unbestritten ist der große historische Einfluß des photoelektrischen Effektes auf den Photonenbegriff; trotzdem läßt sich der Effekt ohne Zuflucht zu einer Quantisierung des elektromagnetischen Feldes erklären. In der Tat kann man das Feld klassisch behandeln, wenn man allein der Materie eine Quantennatur verleiht. Siehe den Aufsatz von W. E. Lamb, Jr. Und M. O. Scully in `Polarisation, Matter and Radiation´, Jubilee Volume in Honor of Alfred Kastler.

Um die Quantelung $\pm h$ des Photonendrehimpulses, den sogen. Photonenspin, ist es nicht besser bestellt. Beth berechnete in seiner oben zitierten Arbeit [7] das für eine bestimmte Lichtmenge der Energie E und der Wellenlänge $\lambda$ zu erwartende Drehmoment in einer $\lambda/2$ – Platte sowohl klassisch als auch nach der Quantentheorie. Nur steckte er hier die gesuchte Drehimpulsquantelung $h$ in Form des Photonenbegriffs mit der Energie $h \cdot \nu$ vorn in die Rechnung hinein. Im Prinzip wird auf Seite 116 der Arbeit zunächst die Anzahl $n$ der Photonen berechnet:

$$n = \frac{E}{h \cdot \nu} \tag{77}$$

Der in der $\lambda/2$ – Platte zu erwartende und auch gemessene Gesamtdrehimpuls $L$ ist dann logischerweise

$$L = n \cdot h \tag{78}$$

oder

$$L = \frac{E}{h \cdot \nu} \cdot h \tag{79}$$

bzw. $\qquad L = \dfrac{E}{\nu}$ *⁾ $\qquad\qquad\qquad\qquad\qquad\qquad\qquad\qquad\qquad\qquad\qquad$ (80)

Zur Berechnung des in diesem Versuch nur gemessenen induzierten Gesamtdrehimpulses L benötigt man also nur die Gleichung (80) mit den Größen E und $\nu$. In die im Prinzip verwendete Gleichung (79) kann man für h jeden beliebigen Wert einsetzen: Er kürzt sich sowieso heraus. Insofern ist das Experiment weder ein Beweis für einen Photonenspin der Größe h noch für die Quantelung eines Drehimpulses von Photonen überhaupt.

---

*⁾ Ersetzt man in Gl. (80) E und $\nu$ durch $E = \tfrac{1}{2} mc^2$ und $\nu = \dfrac{c}{\lambda}$ so erkennt man die Entstehung des Drehimpulses in der $\lambda/2$ − Platte als Produkt aus Translationsimpuls und Wirkungsabstand: $L = \dfrac{1}{2}\dfrac{mc^2}{\nu} = m \cdot c \cdot \dfrac{\lambda}{2}$.